\documentclass[twocolumn]{aastex631}

\usepackage{graphicx}
\usepackage{apjfonts}
\usepackage{hyperref}
\usepackage{subfigure}

\submitjournal{ApJ}

\shorttitle{Global impact of emerging internetwork fields on the low solar atmosphere} 

\shortauthors{Go\v{s}i\'{c} et al.}

\begin{document}
	
\title{Global impact of emerging internetwork fields on the low solar atmosphere}

\correspondingauthor{M.~Go\v{s}i\'{c}}
\email{gosic@baeri.org}

\author[0000-0002-5879-4371]{M.~Go\v{s}i\'{c}}
\affil{Lockheed Martin Solar and Astrophysics Laboratory, Palo Alto, CA 94304, USA}
\affil{Bay Area Environmental Research Institute, Moffett Field, CA 94035, USA}

\author[0000-0002-8370-952X]{B.~De Pontieu}
\affil{Lockheed Martin Solar and Astrophysics Laboratory, Palo Alto, CA 94304, USA}
\affil{Institute of Theoretical Astrophysics, University of Oslo, P.O. Box 1029 Blindern, NO-0315 Oslo, Norway}
\affil{Rosseland Centre for Solar Physics, University of Oslo, P.O. Box 1029 Blindern, NO-0315 Oslo, Norway}

\author[0000-0002-3234-3070]{A.~Sainz Dalda}
\affil{Lockheed Martin Solar and Astrophysics Laboratory, Palo Alto, CA 94304, USA}
\affil{Bay Area Environmental Research Institute, Moffett Field, CA 94035, USA}

\begin{abstract}
Small-scale, newly emerging internetwork (IN) magnetic fields are considered a viable source of energy and mass for the solar chromosphere and possibly the corona. Multiple studies show that single events of flux emergence can indeed locally heat the low solar atmosphere through interactions of the upward propagating magnetic loops and the preexisting ambient field lines. However, the global impact of the newly emerging IN fields on the solar atmosphere is still unknown. In this paper we study the spatio-temporal evolution of IN bipolar flux features and analyze their impact on the energetics and dynamics of the quiet Sun atmosphere. We use high resolution, multi-wavelength, coordinated observations obtained with the Interface Region Imaging Spectrograph (IRIS), Hinode and the Solar Dynamics Observatory (SDO) to identify emerging IN magnetic fields and follow their evolution. Our observational results suggest that only the largest IN bipoles are capable of heating locally the low solar atmosphere, while the global contribution of these bipoles appears to be marginal. However, the total number of bipoles detected and their impact estimated in this work is limited by the sensitivity level, spatial resolution, and duration of our observations. To detect smaller and weaker IN fields that would maintain the basal flux, and examine their contribution to the chromospheric heating, we will need higher resolution, higher sensitivity and longer time series obtained with current and next-generation ground- and space-based telescopes.
\end{abstract}

\keywords{Sun: magnetic field -- Sun: photosphere -- Sun: chromosphere -- Sun: transition region}

\section{Introduction}
\label{sec1}

Internetwork (IN) magnetic fields are dynamic magnetic structures that populate the interior of supergranular cells \citep{LivingstonHarvey, Smithson1975}. They are spread all over the Sun \citep{Wang}, maintain the photospheric network (NW) \citep{Gosicetal2014}, and may hold a significant fraction of the total magnetic energy stored at the solar surface \citep{TrujilloBueno2004, 2021ApJ...915L..20T}. For these reasons IN fields are considered to be the main building blocks of the quiet Sun (QS) magnetism (see \citealt{BellotRubioOrozcoSuarez2019} for a review). 

Recent Hinode observations showed that IN fields mainly appear in the form of magnetic bipoles in the photosphere \citep{2022ApJ...925..188G}, likely generated by small-scale surface dynamo \citep{Rempel2014}, but other scenarios are also possible \citep{2011ApJ...729..136P, 2012ApJ...755..175M, 2023ApJ...944...95T}. According to some numerical models \citep{Isobeetal2008, Amarietal2015, MorenoInsertisetal2018}, and observations \citep[e.g.,][]{MartinezGonzalezBellotRubio2009, MartinezGonzalezetal2010, 2021ApJ...911...41G}, these fields may upon appearance in the photosphere rise through the lower atmosphere, and locally heat the chromosphere and transition region. 

Considering the magnetic and energy budget of IN fields, it is important to determine the global contribution of the emerging IN fields to the dynamics and energetics of the chromosphere and the atmospheric layers above. This open question has not yet been addressed in detail, using high resolution observations that simultaneously cover the solar atmosphere from the photospheric to coronal heights. The main reason for this was the lack of suitable observations and the need for sophisticated analysis that allows to identify footpoints of magnetic loops, and determine their history in a reliable way. Such observations at the photospheric level are provided by the Narrowband Filter Imager \citep[NFI;][]{Tsuneta} aboard the Hinode satellite \citep{2007SoPh..243....3K}, and at the chromospheric/transition region and coronal levels by the Interface Region Imaging Spectrograph \citep[IRIS;][]{DePontieuetal2014} and the Atmospheric Imaging Assembly \citep[AIA;][]{Lemenetal2012} onboard the Solar Dynamics Observatory \citep[SDO;][]{Pesnelletal2012}. Furthermore, to understand the impact of newly emerging IN fields on the chromospheric energy balance, one would need to determine the thermodynamic properties from chromospheric lines, considering the non-local thermodynamic equilibrium (non-LTE) radiative transfer. Diagnosing chromospheric conditions requires inversion codes that, due to the physics necessary to be implemented (non-LTE, partial frequency redistribution and atom models), are typically slow and difficult to use. In this work we will take advantage of a new approach that solves these issues through a combination of machine learning and classical inversion techniques to speed up and facilitate the recovery of thermodynamical information from the solar spectra \citep[e.g.,][]{SainzDaldaetal2019}.

The work presented in this paper builds upon our previous efforts to gain a better understanding of small-scale flux emergence \citep{2021ApJ...911...41G}. In this study, we carry out a statistical analysis to determine the global impact of newly emerging IN fields on the low solar atmosphere. We address this open question by employing coordinated, multi-wavelength observations from IRIS, Hinode and SDO. These instruments allow us to study the spatio-temporal evolution of the QS fields at high spatial, spectral, and temporal resolution, while observing the solar atmosphere from the photosphere up to the transition region and corona. 

The observations used in this paper are described in Section \ref{sec2}. The 
identification, classification and tracking of IN bipolar flux features is explained in Section \ref{sec3}. Section \ref{sec4} provides the results, while conclusions are given in Section \ref{sec5}.

\section{Observations and data processing}
\label{sec2}

Our observations are obtained on 2013 September 23. IRIS measurements start at 07:09:49~UT and end at 12:05:37~UT. Hinode measurements cover this interval from 08:04:38~UT to 10:59:36~UT. The observations show the spatio-temporal evolution of a QS region at the disk center.

IRIS data set is a medium-sit-and-stare raster, taking spectra in the near ultraviolet (NUV) band\footnote{IRIS also takes spectra in two far ultraviolet domains, which were not used in this paper.} in the wavelength range from 2790~\AA\ to 2835~\AA\/. The NUV spectroscopic measurements sample the solar atmosphere from the photosphere to the upper chromosphere. The spectra are recorded every 5 seconds along a slit length of $60$\arcsec. Slit-jaw images (pixel size is $0\farcs16$) were taken using the \ion{C}{2} 1330~\AA\ (SJI 1330), \ion{Si}{4} 1400~\AA\ (SJI 1400), \ion{Mg}{2} k 2796~\AA\ (SJI 2796), and \ion{Mg}{2} h wing at 2832~\AA\ (SJI 2832) filters, compensating for the solar rotation. The cadence of the slit-jaw images are 18~s, 15~s, 15~s, and 89~s, respectively. The IRIS data were corrected for dark current, flat-field, geometric distortion, and scattered light \citep{2018SoPh..293..149W}.

Using the IRIS$^{2}$ inversion code\footnote{The IRIS$^{2}$ code is publicly available in the IRIS tree of SolarSoft. For more details about the code see \url{https://iris.lmsal.com/iris2}.} \citep{SainzDaldaetal2019} we derived the thermodynamical properties of the observed QS atmosphere as a function of the optical depth. The code employs the k-means clustering method to build a database of the representative IRIS \ion{Mg}{2} h and k spectral profiles (RP) and their corresponding atmospheric models. These RPs were inverted with the STiC code\footnote{STiC is publicly available to the community from the author’s repository: \url{https://github.com/jaimedelacruz/stic}.} \citep{delaCruzRodriguezetal2016, delaCruzRodriguezetal2019}. For each observed \ion{Mg}{2} h and k pair, IRIS$^{2}$ assigns the model atmosphere resulting from the inversion of the closest RP to the observed profiles.

The NFI was employed in shutterless mode to obtain the full Stokes vector at 2 wavelength positions around the core of the photospheric \ion{Fe}{1} 5250~\AA\ line. These observations provide circular and linear polarization maps, showing photospheric activity of the vertical (loop footpoints) and horizontal (loop tops) components of magnetic fields, respectively. After the data reduction process and co-alignment of Hinode and IRIS observations, the effective field of view (FOV) was reduced to $35\arcsec \times 60\arcsec$, which is sufficient to capture the evolution of at least two supergranular cells for two hours and 40 minutes at a cadence of $\sim60$~s. This allows us to track the temporal evolution of IN fields in a magnetogram sequence that considerably exceeds the mean lifetime of IN magnetic structures on granular scales. Magnetograms $M$ were calculated using the Stokes $I$ and $V$ filtergrams:

\begin{equation}
\label{mag_eq}
M=\frac{1}{2}\left(
\frac{V_{\text{blue}}}{I_{\text{blue}}}-\frac{V_{\text{red}}}{I_{\text{red}}}
\right),
\end{equation}

\noindent where ``blue'' indicates the measurements in the blue wing of the line and ``red'' in the red wing. The linear polarization maps $\rm LP$ are computed as ${\rm LP}= \sum_{i=1}^{2} \sqrt{Q(\lambda_i)^2 + U(\lambda_i)^2}/I(\lambda_i)/2$. All magnetogram and LP maps were smoothed using a $3\times3$ Gaussian-type spatial kernel to reduce the noise, and the five-minute oscillations were removed from the maps by applying a subsonic filter \citep{1989ApJ...336..475T, 1992A&A...256..652S}.

\begin{figure*}[!t]
	\centering \includegraphics[width=1\textwidth]{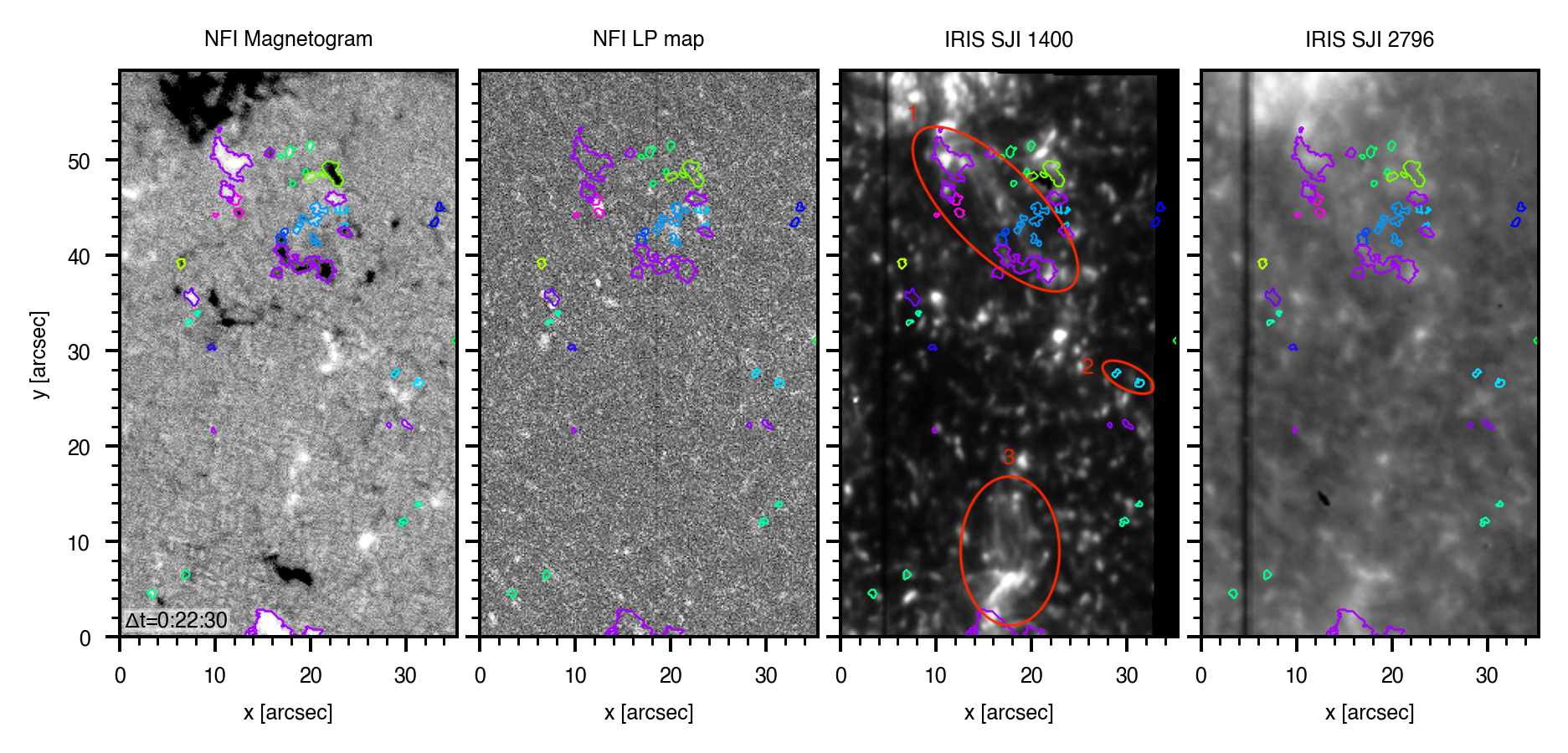} 
	\caption{\textit{From left to right:} Hinode/NFI magnetograms and linear polarization maps, IRIS SJI 1400 and SJI 2796 slit-jaw images. The detected bipoles are enclosed with contours of different colors. Regions 1, 2, and 3 (red ellipses) show the largest emerging cluster of magnetic elements, one small-scale internetwork loop, and a network patch originating in a previously emerged internetwork bipole.\newline
	({\em An animation of this figure is available and runs from $\Delta t$=0:00:00 to $\Delta t$=2:36:45.})}
	\label{fig1}
\end{figure*}

We also made use of the Helioseismic and Magnetic Imager \citep[HMI;][]{Scherreretal2012} and AIA observations. This allows us to determine the evolution of the observed QS region before the IRIS and Hinode observations started and to examine emission at coronal heights.

The alignment of the datasets was carried out by compensating for solar rotation and scaling all images to match the IRIS pixel size. All IRIS, Hinode and SDO sequences were interpolated in time applying the nearest neighbor method of interpolation to match the cadence of the SJI 2796 images (15~s). Images are then aligned comparing prominent NW features and bright points in SJI 2832 images with the Hinode intensity filtergrams and AIA 1600 \AA\/ channel.

\section{Identification and tracking of internetwork bipoles}
\label{sec3}

To understand how bipolar IN magnetic structures globally affect the low solar atmosphere, we detected and tracked all IN bipoles (loops and clusters) visible in the FOV. We first automatically identified and tracked these bipoles in the photosphere using Hinode observations, and then examined their impact on the chromospheric activity using IRIS and SDO observations. Below we describe in detail our methods to identify and track IN magnetic bipoles as they emerge through the solar atmosphere.

\subsection{Emergence of internetwork bipoles in the photosphere}

To detect IN magnetic bipoles (loops and clusters) and separate them from the unipolar fields (isolated flux concentrations), we first identified all individual magnetic features in the magnetograms and LP maps. We consider loops to represent two circular polarization patches (positive and negative polarity footpoints) moving away from each other, while flux clusters consist of two or more patches that emerge within a short time interval in a relatively small area.

Using the YAFTA code\footnote{YAFTA (Yet Another Feature Tracking Algorithm) is an automatic tracking code and can be downloaded from the author's website at \url{http://solarmuri.ssl.berkeley.edu/~welsch/public/software/YAFTA/}.} and the downhill identification method \citep[]{WelschLongcope}, we automatically tracked all the detected flux patches to determine their spatio-temporal evolution. This process includes identifications of all merging, fragmentation and cancellation events that magnetic patches may undergo during their lifetimes. In this way, we can determine the history of every detected magnetic feature. Real features in the magnetograms are separated from the background signal by setting a flux density threshold to $2\sigma$ (10~Mx~cm$^{-2}$). This allows us to detect more faint magnetic elements, considering all of them to be real if their minimum size is at least 4 pixels and they live two frames or more. Magnetic features that appear and disappear in situ and are visible in only one frame are discarded because they may represent intrinsic flux fluctuations around the threshold level. 

The appearance of footpoints in the photosphere is preceded by LP signal between them. Therefore, magnetic bipoles are identified by searching for all LP signals (loop tops) that are followed by pairs of opposite-polarity flux features that appear in situ (loop footpoints). To be selected, these footpoints have to appear within 6 minutes after the first one becomes visible in a magnetogram, and they have to move away from each other. Although clusters bring numerous magnetic patches to the solar surface, they follow the same pattern of the spatio-temporal evolution as loops, i.e., flux features move away from each other with respect to their common center of appearance. Usually, there are multiple LP patches within clusters. The tracking and identification of IN bipoles in this work is similar to the method described in \cite{2022ApJ...925..188G}, the difference being that here we use the NFI LP maps instead of extrapolations of the magnetic field lines to identify the loop tops.

For the strongest flux patches visible in the first frame, we cannot determine their history from the NFI magnetograms. Thus, we used HMI data to determine if they appear as bipolar structures. This is important because the strongest magnetic elements may considerably impact the low solar atmosphere. Since HMI is not sensitive to the weakest IN patches, we classified as bipoles only those flux patches that are clearly resolved and appear in situ, following the expected pattern of flux emergence.

\subsection{Chromospheric response to the emerging internetwork fields}
\label{tracking_chromosphere}

In this section, we describe how a significant level of the chromospheric activity above flux emerging regions is identified and estimated. We carried out both manual and automatic identification to increase the reliability of our results. To detect  any chromospheric activity related to the newly emerging IN fields, we analyzed SJI 1400 and 2796 images. We assumed that the emerging IN fields that could affect the chromosphere generate bright features in SJI 1400 and possibly in SJI 2796 filtergrams. Limitations of our methods to detect chromospheric activity are also discussed.

\subsubsection{Manual identification}

We manually inspected all the detected IN bipoles by visually examining the chromospheric activity in SJI 1400 and SJI 2796 images above the emerging field regions. This includes the identification of all bright features that appear above and between the photospheric footpoints in those images. In this way, we can identify possible chromospheric heating imprints in the regions where there is already an increased chromospheric activity, for example, when an IN bipole emerges next to a strong NW flux patch. Visual inspection allows us to detect loop-like structures, bright grains, and jets forming due to interactions of the emerging and the preexisting magnetic fields.
	
Emerging IN loops crossed by the IRIS slit were detected by visual inspection. In total, we found seven such bipoles, and their impact on the chromosphere is estimated by examining the temperature map obtained through IRIS$^2$ inversions of the \ion{Mg}{2} h and k spectral lines.

Since manual identification may be slow and subjective, we developed an automatic method to search for the chromospheric activity above emerging IN fields. We describe this method in the following.

\subsubsection{Automatic identification}

In our data, the chromospheric activity above the flux emerging regions depend on magnetic fields rising through the solar atmosphere, but also on natural variability in the SJI 1400 signal. We will assume here that the natural SJI 1400 variability includes all possible mechanisms that may contribute to the chromospheric activity, but are not related to the emerging fields. To distinguish between the contributions from the emerging fields and natural/basal SJI signal variability, we analyzed SJI 1400 intensities in QS areas considering all the pixels with the signal above a threshold level of 30 counts per pixel. The SJI 1400 filter is sensitive to emission from the transition region \ion{Si}{4} 1394/1403~\AA\ lines and continuum formed in the upper photosphere/lower chromosphere. Therefore, SJI 1400 images of the QS are dominated by short-living (less then 2 minutes), periodic (2 to 4 minutes), point-like chromospheric grains (for an extensive review see \citeauthor{1991SoPh..134...15R} \citeyear{1991SoPh..134...15R}; see also \citeauthor{MartinezSykoraetal2015} \citeyear{MartinezSykoraetal2015}; \citeauthor{2023ApJ...955L..40M} \citeyear{2023ApJ...955L..40M}, and references therein). Because of this, we used in our analysis both non-filtered SJI 1400 images and also filtered images from which the short-living, transient SJI 1400 brightenings are removed.

After identifying regions where we do not detect footpoints of magnetic bipoles, i.e., where there are no emerging IN fields in the magnetograms, we randomly selected 161 of them (same as the number of the detected bipolar structures). The period for which a given identified region remains empty from bipolar footpoints is used to create two sub-sequences of different lengths: one acting as a pre-emerging phase (lasting between $15$ seconds and $5$ minutes), and the second representing a fake emerging flux phase (varied from $5$ to $16$ minutes, which is the mean lifetime of the detected footpoints in our observations). Finally, we calculated the average SJI 1400 intensities in the two sub-sequences and determined their ratios.

For the resulting activity ratio histograms we calculated single Gaussian fits and found that the activity ratios in filtered and non-filtered images are slightly different. Generally, non-filtered images have broader distributions with an average mean value of $1.02$ and $\sigma=0.15$. On the other hand, filtered SJI 1400 images result in narrower distributions with an average mean value of $1.01$ and $\sigma=0.09$.

\begin{figure}[!t]
	\centering
	\resizebox{1\hsize}{!}{\includegraphics[width=1.0\textwidth]{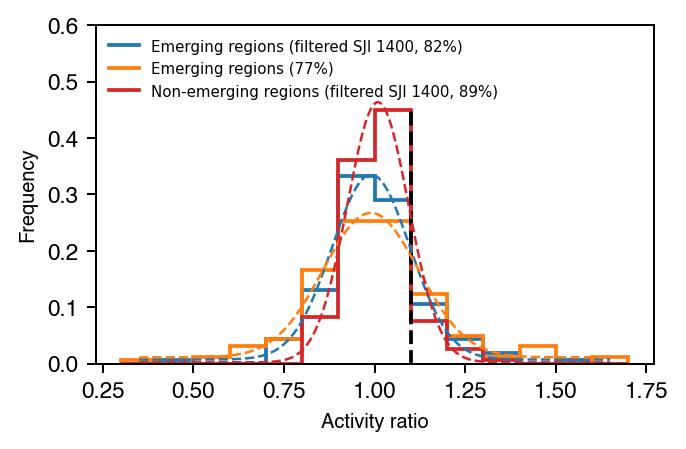}}
	\caption{Activity ratios derived from non-filtered (orange) and filtered (blue) IRIS SJI 1400 images. The ratio equal to or below $1.1$ (vertical black dashed line) indicates that the activity above a given emerging event did not increase with time. The red solid line shows the activity ratio distribution for non-emerging regions in filtered SJI 1400 images. The dashed lines represent single Gaussian fits to the corresponding histograms. Numbers in the parentheses indicate percentages of the detected bipoles with activity ratios below $1.1$.}
	\label{fig4}
\end{figure}

If we consider the filtered SJI images from which the transient SJI 1400 brightenings are removed, at the $1\sigma$ level of significance, ratios below $1.1$ would represent regions where the chromospheric activity did not increase over the observed time interval. In our data sets, this translates into $82\%$ (in filtered images) and $77\%$ (in non-filtered images) of the detected bipoles that did not generate higher chromospheric activity during their lifetimes. The corresponding histograms (solid lines) and Gaussian fits (dashed lines) are shown in Figure~\ref{fig4} with the blue and orange curves, respectively. The red solid (histogram) and dashed (Gaussian fit) lines represent an example of the activity ratio distributions for non-emerging regions using filtered SJI 1400 images.

\begin{figure*}[!t]
	\centering \includegraphics[width=0.99\textwidth]{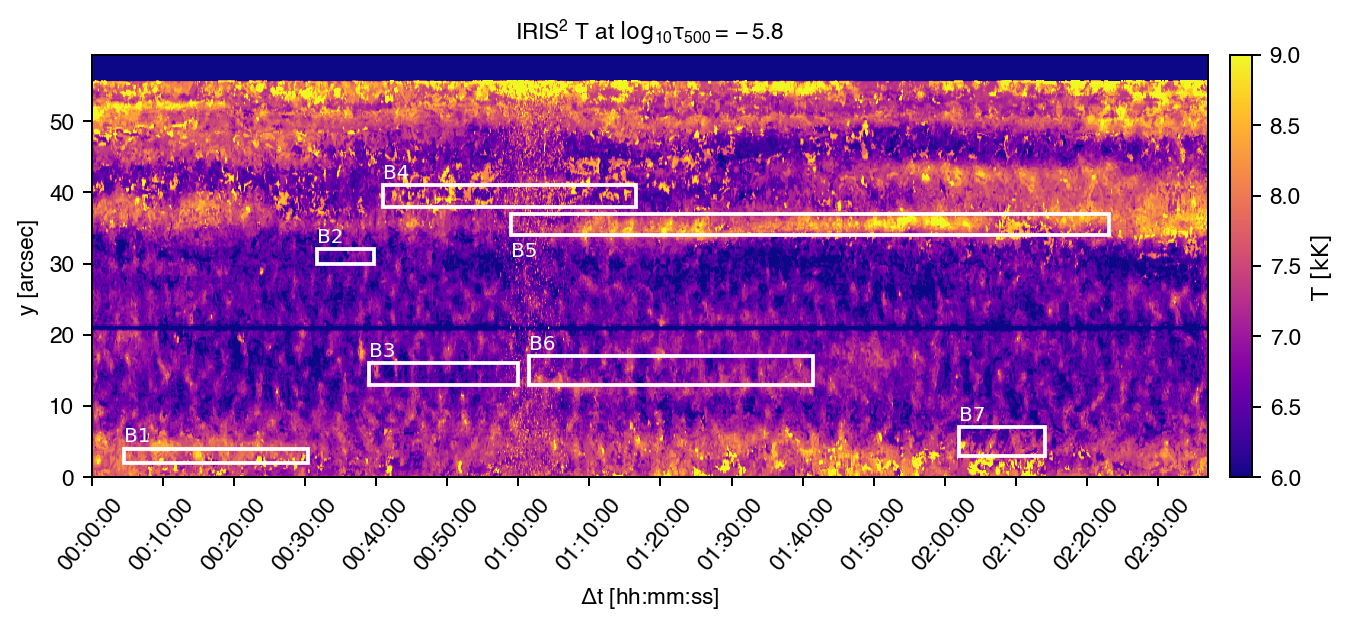} 
	\caption{Temperature spatio-temporal map from the IRIS$^{2}$ inversions at $\log_{10}\tau_{500}=-5.8$. The white boxes indicate locations and times when the emerging IN bipoles were under the IRIS slit.}
	\label{fig2}
\end{figure*}

\subsubsection{Uncertainties}

The manual identification (i.e., visual inspection) of the chromospheric response to the emerging fields allows us to examine in detail magnetic loops that may perturb the chromosphere. In this way, we can examine the chromospheric activity among and above the footpoints of the emerging loops, including not only the overlapping brightenings but also eruptions that may result from reconnections of the emerging and the preexisting fields. While such an approach is in general very precise, it is hardly repeatable and may suffer from subjectivity. Therefore, we also developed an automatic detection of the chromospheric response to the newly emerging loops. However, while automatic identification and tracking is repeatable, it is not without its limitations.

The downside of the automatic detection of the chromospheric activity is that bipolar features may appear in highly dynamic environments, such as near NW regions that are generally very active. In such cases, it can be challenging to distinguish whether the chromospheric activity is being driven by the newly emerging loops or preexisting ambient fields. If the activity above the flux emerging regions was higher before the newly IN bipoles emerged, our method would classify such bipoles as those that do not contribute to the chromospheric energetics and dynamics.

As explained in the previous section, we used an activity ratio of $1.1$ as a threshold. Varying the fake emerging flux phase from one frame ($15$~seconds) to $16$ minutes, the activity ratio changes between $1.1$ and $1.2$. However, the latter turned out to be very restrictive in our tests, missing to detect numerous SJI 1400 loops and brightenings and resulting in only up to $10\%$ of bipoles that contribute to the chromospheric heating.

It is worth noticing that the values below $1.1$ may not necessarily mean that the emerging loops do not heat locally the chromosphere. It is possible that the bipoles appearing within already active environments (e.g., next to the NW fields), heat the lower solar atmosphere, but their contribution cannot be distinguished from an ongoing heating.
	
Another important aspect is that the activity ratio is not entirely reliable as it depends on the regions where IN loops emerge. Based on the estimated basal chromospheric activity in our SJI 1400 images, we have determined that a $10\%$ increase in activity above an emerging region is needed to trigger positive identification of an increased chromospheric activity. This increase can occur through a gradual, steady increase of the signal over the course of the bipole's lifetime, or through sudden and strong brightenings that would be visible in at least a couple of frames. In a low-activity, non-emerging regions, this increase is easily achieved as SJI 1400 brightenings can generate signals of several tens to several hundreds of counts above the background level. However, in an already active region where the SJI 1400 signals may be very strong, obtaining a $10\%$ higher activity could be difficult. This is the reason why our automatic identification may fail detecting contributions to the chromospheric heating from IN bipoles emerging close to stronger, network-like magnetic fields.

\vspace{1em}
\section{Results}
\label{sec4}
\vspace{1em}

Using HMI and Hinode/NFI magnetograms we identified and tracked the spatio-temporal evolution of individual magnetic elements representing footpoints of $161$ bipolar structures (IN loops and clusters of magnetic elements). This translates into an emergence rate of $\sim0.038$~bipoles per hour and arcsec$^{2}$, which is in agreement with the results reported by \cite{2022ApJ...925..188G}. If we take into account the total area occupied by the footpoints, only $2\%$ of the available FOV at any given time is covered by bipolar IN fields. Note that this ratio depends on spatial resolution, magnetic sensitivity, and intrinsic fluctuations of the total instantaneous unsigned IN flux \citep[e.g.,][]{2022ApJ...925..188G}. 

The two largest bipoles whose footpoints are visible in the first NFI magnetogram are identified employing HMI data. They are marked in Figure~\ref{fig1} with violet contours inside the encircled region 1 (red ellipse) and at the bottom of the FOV at $(x, y) = (16\arcsec, 0\arcsec)$. By the time Hinode/NFI started to observe, most of the footpoints of the two clusters either transformed into NW features or canceled with the opposite polarity NW elements. The tracking results can be evaluated with the animation accompanying Figure~\ref{fig1}, which shows all the magnetic bipoles (left panel) detected in our Hinode/NFI magnetograms. Flux patches belonging to the same bipole have the same colors. The corresponding contours are overplotted on the LP maps, SJI 1400 and SJI 2796 images, from left to right, respectively.

\begin{figure*}[!t]
	\centering \includegraphics[width=0.99\textwidth]{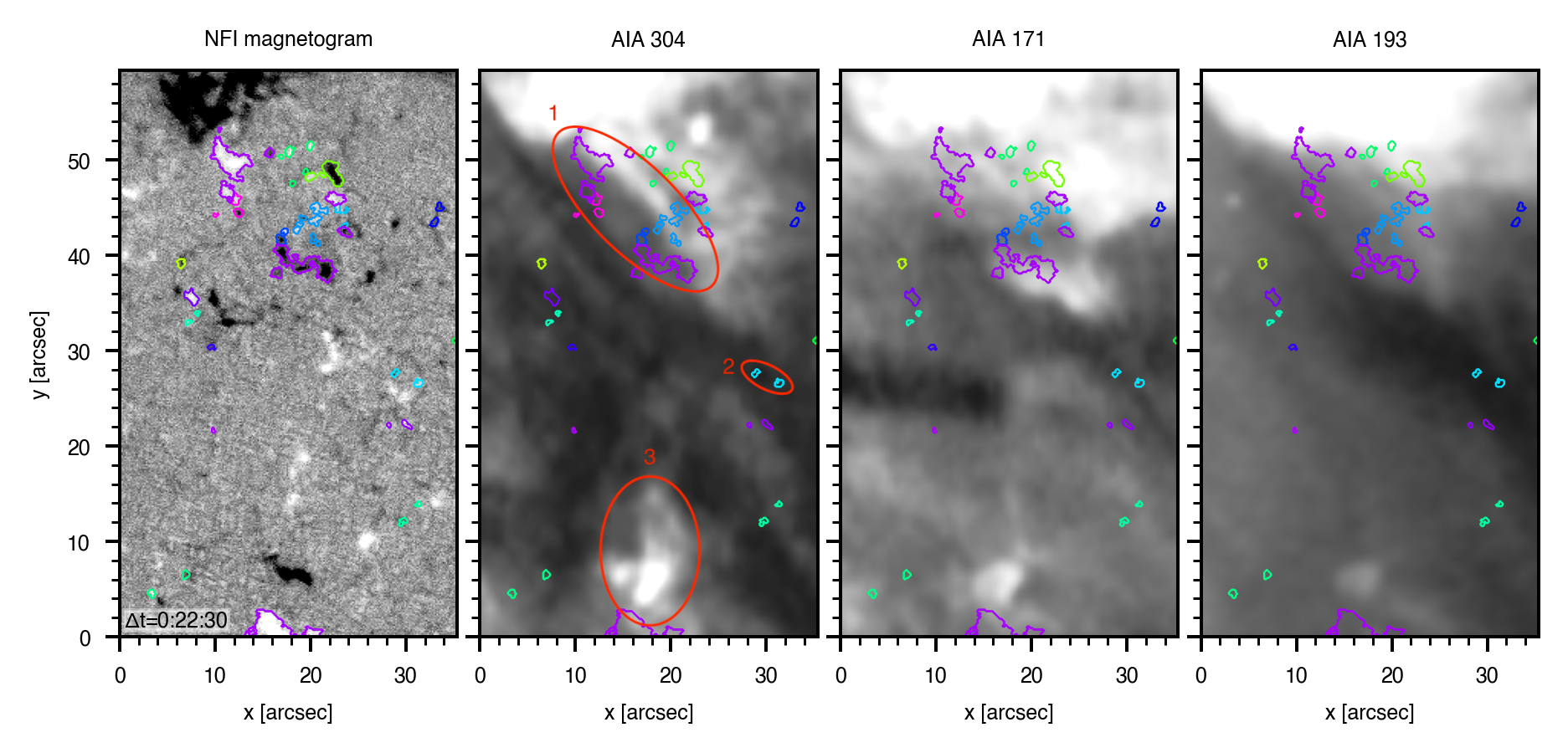} 
	\caption{\textit{From left to right:} Hinode/NFI magnetogram, AIA 304 \AA\/, AIA 171 \AA\/, and AIA 193 \AA\/ images. The detected IN bipoles are enclosed with contours having different colors. The red ellipses encloses the same regions as the ones shown in Figure~\ref{fig1}.\newline
		({\em An animation of this figure is available and runs from $\Delta t$=0:00:00 to $\Delta t$=2:36:45.})}
	\label{fig3}
\end{figure*}

As can be seen from Figure~\ref{fig1} and the accompanying movie, the strongest emission in both filtergram sequences is co-spatial with strong magnetic elements, i.e., large clusters and the positive- and negative- polarity NW elements centered at $(x, y) = (16\arcsec, 0\arcsec)$ and $(x, y) = (10\arcsec, 56\arcsec)$, respectively. The rest of the FOV is overwhelmed by smaller bright features. By visual inspection of IRIS SJI features above the detected bipoles, we determined that most of the bipoles are either embedded in regions with already ongoing activities in the chromosphere or the overlapping SJI 1400 brightenings above them do not seem to be different from the background activity. Such an example is the loop inside region 2 shown in the SJI 1400 panel in Figure~\ref{fig1} and Figure~\ref{fig3}. As a reference, this loop has an activity ratio of $1.02$. In contrast, the positive polarity magnetic element inside region 3 clearly impacts locally the chromosphere through interactions with the surrounding opposite-polarity flux features. Numerous dynamic and episodic bright loops and grains can be seen around this footpoint, probably energized by reconnection of the magnetic field lines of the footpoint and the surrounding flux patches. The cluster within region 1 exhibits a temporal and spatial evolution similar to the cluster described in \cite{2021ApJ...911...41G}. The footpoints expanded with time and started interacting with the negative polarity NW patches in the north. Eventually the region produced a surge-like event around $\Delta t$=00:25:30, (onset at $\Delta t$=00:22:30), which is expected to be observed when new and preexisting fields reconnect \citep{Guglielminoetal2018, NobregaSiverioetal2017}. In total, we find that $28\%$ of the detected loops contribute to the chromospheric heating when manual identification is used.

Employing the previously described automatic identification method, and bearing in mind its limitations, the total number of IN bipoles heating the low solar atmospheres is estimated to be $23\%$ in non-filtered and $18\%$ in filtered SJI 1400 images. We show in Figure~\ref{fig4} the resulting distributions of the ratios of the average SJI 1400 signals in non-filtered (orange) and filtered (blue) images just before the footpoints appear and during their lifetimes.

Figure~\ref{fig2} shows the temperature map derived from IRIS$^{2}$ inversions. The IRIS slit covered seven emerging bipoles. Five of them are embedded in the background activity and do not perturb the chromosphere. They do not produce any excess emission neither in the IRIS NUV nor FUV spectral lines. The bipole labeled as B4, is co-spatial with increased temperature, but this is likely due to cancellation with the opposite polarity magnetic features in its vicinity \citep[e.g.,][]{Gosicetal2018}. Only a few bipoles can be associated with the chromospheric activity. For example, the negative polarity footpoint (B5), clearly shows an increase of the chromospheric temperature. This magnetic element eventually becomes an NW element. Bipole B1 emerges next to an ongoing cancellation event (hence a higher temperature before the bipole emerged), with which the positive footpoint starts interacting and eventually completely disappears. This cancellation maintained an increased temperature in that region for the next 26 minutes. Another intriguing event is B2 loop that shows a slight temperature increase towards the end of the observed temporal window (white box). This is probably the result of an upward propagating wave because we do not see any activity above the footpoints in the filtered IRIS slit-jaw images. In addition, this is a small, short-lived loop (8 minutes), so it is unlikely that in such a short time this loop can reach the chromosphere.

Very limited activity in the lower solar atmosphere within the observed QS region is also apparent in the AIA filtergrams displayed in Fig~\ref{fig3}. The AIA 304~\AA\/, 171~\AA\/ and 193~\AA\/ channels show the chromospheric (304~\AA\/) and coronal (171~\AA\/ and 193~\AA\/) activity inside regions 1 and 3. The rest of the FOV looks very quiet with some long loops extended across the FOV that originate in an active region in the north (not visible in the Hinode and IRIS observations).

\section{Conclusions}
\label{sec5}

In this work we used Hinode/NFI, IRIS and SDO/AIA observations to detect newly emerging IN bipoles in the solar photosphere and estimate the global and direct contribution of emerging fields on the chromospheric dynamics and energetics. Our results suggest that the majority of IN bipoles (at least $72\%$) may not have enough magnetic buoyancy nor live long enough to rise through the solar atmosphere and directly affect the solar chromosphere and beyond (we detected only three such strong bipoles). This result should be understood as a minimum -- more active QS regions may generate stronger emerging fields capable of rising through the solar atmosphere. Also, our statistics could have been different had we identified more bipoles under the slit, which will be possible in future with multislit instrument such as MUSE \citep{2020ApJ...888....3D}. In the meantime, if we consider only the bipoles under the slit, despite of the small sample, then $40\%$ of the loops may heat the chromosphere either directly through reconnection with the overlying magnetic fields or through cancellation of the footpoints with the surrounding flux patches after the bipoles lose their coherence by merging and fragmentation processes.

Based on our observations, only the strongest three detected bipoles noticeably produced local temperature increase in the chromosphere. They are capable of generating surge-like phenomena through reconnection of their magnetic field lines with the preexisting fields. We conclude that newly emerging IN bipoles, at the sensitivity levels and spatial resolution of Hinode/NFI magnetograms, cannot globally maintain the chromospheric heating directly through interaction with the ambient overlying magnetic fields. We either do not see a lot of evidence of heating, except for larger events, or the large events are too sporadic in space and time to considerably support the chromospheric heating. It would be interesting to study longer-duration events to increase the statistical sample that can be studied under the slit. We also note that our analysis has been focused on detecting changes in emission or chromospheric temperature as a result of the detected emergence of IN magnetic elements. We did not investigate a scenario in which undetected or undetectable IN elements may continuously and ubiquitously occur in the photosphere and  lead to a steady heating of the atmosphere or continuous background emission. The results presented in this paper do not exclude a possibility that the footpoints of IN bipoles may possibly contribute to the chromospheric heating indirectly through other mechanisms such as magneto-acoustic waves and shocks, braiding of the magnetic field lines and swirls. 

To better understand the smallest and weakest QS fields, we will need long-duration observations with higher spatial resolution and sensitivity. Such observations could be obtained with the Daniel K. Inouye Solar Telescope \citep[DKIST;][]{Elmoreetal2014}, and from space with the Solar Orbiter's Polarimetric and Helioseismic Imager \citep{2020A&A...642A..11S}. These instruments can detect and resolve fields that are not accessible to the currently available telescopes, but they may be continuously emerging and contributing to heating of the lower atmosphere.

A key aspect of this issue is also to study which processes determine whether emerging IN fields rise through the solar atmosphere and transfer mass and energy. This will be investigated in detail in our future work using radiative MHD Bifrost simulations (\citeauthor{2011A&A...531A.154G} \citeyear{2011A&A...531A.154G}; see also \citeauthor{2023ApJ...944..131H} \citeyear{2023ApJ...944..131H}).
 
\begin{acknowledgments}
The authors would like to thank the anonymous referee who provided valuable feedback and suggestions to improve the manuscript. MG is supported by NASA contracts NNG09FA40C (IRIS) and NNM07AA01C (Solar-B (Hinode) Focal Plane Package Phase E). BDP and ASD are supported by NASA contract NNG09FA40C (IRIS). We acknowledge the use of IRIS, Hinode and SDO/AIA/HMI data. IRIS is a NASA Small Explorer Mission developed and operated by LMSAL with mission operations executed at NASA Ames Research Center and major contributions to downlink communications funded by ESA and the Norwegian Space Centre. The Hinode data used here were acquired in the framework of the Hinode Operation Plan 243 {\em ``Effects of Quiet Sun Weak Fields on the Chromosphere and Transition Region.''}. Hinode is a Japanese mission developed and launched by ISAS/JAXA, with NAOJ as a domestic partner and NASA and STFC (UK) as international partners. It is operated by these agencies in co-operation with ESA and NSC (Norway). AIA is an instrument on board the Solar Dynamics Observatory, a mission for NASA’s Living With a Star program. This research has made use of NASA’s Astrophysics Data System.
\end{acknowledgments}

\end{document}